\documentclass[aps,pra,twocolumn,superscriptaddress,showpacs,floatfix]{revtex4-2}
\usepackage[dvipdfmx]{graphicx}
\usepackage[dvipdfmx]{color}
\usepackage{graphicx}
\usepackage{bm}
\usepackage{color}
\usepackage{amsmath}
\usepackage{comment}
\begin{document}

\title{Multi-state interferometric measurement of nonlinear AC Stark shift}
\author{Junnosuke Takai}
\email{takai@qo.phys.gakushuin.ac.jp}
\affiliation{Department of Physics, 
Gakushuin University, Tokyo 171-8588, Japan}
\author{Kosuke Shibata}
\affiliation{Department of Physics, 
Gakushuin University, Tokyo 171-8588, Japan}
\author{Naota Sekiguchi}
\affiliation{Department of Electrical and Electronic Engineering, 
Tokyo Institute of Technology, Tokyo 152-8552, Japan}
\author{Takuya Hirano}
\affiliation{Department of Physics, 
Gakushuin University, Tokyo 171-8588, Japan}
\date{\today}

\begin{abstract}
We demonstrate measurement of quadratic AC Stark shifts between Zeeman sublevels in an $^{87}$Rb Bose--Einstein condensate using a multi-state atomic interferometer. 
The interferometer can detect a quadratic shift without being affected by relatively large state-independent shifts, thereby improving the measurement precision.
We measure quadratic shifts in the total spin $F = 2$ state due to the light being near-resonant to the D$_1$ line. 
The agreement between the measured and theoretical detuning dependences of the quadratic shifts 
confirms the validity of the measurement. 
We also present results on the suppression of nonlinear spin evolution using near-resonant dual-color light pulses with opposite quadratic shifts. 
\end{abstract}

%\pacs{}

\maketitle
\section{Introduction}

Atom--field interactions often cause shifts in atomic energy levels, 
referred to as AC Stark shifts in the semi-classical treatment.
% The interaction between light and atom is basic tools to realize some quantum technologies in the field of atomic physics.
% This interaction sometimes causes shifts on the atomic energy level, and the AC Stark interaction is common treatment in the case that atom is affected by the classical oscillating electric fields. 
% The AC Stark interaction between light and atoms is used in various experiments. 
These shifts often depend on the atomic spin state. 
Linear and quadratic energy shifts (light shifts) of Zeeman sublevels can arise from 
spherical tensor operators of, respectively, rank-1 (vector) and 2 (tensor) in the irreducible decomposition of the interaction Hamiltonian \cite{Geremia2006, Deutsch2010}. 
Quadratic shifts enable advanced quantum state manipulation such as dynamical spin control \cite{Chaudhury2007} and nuclear--electronic spin entanglement \cite{fernholz2008spin}. 
Quadratic shifts have recently been used to generate the Schr\"{o}dinger's kitten state in cold Dy atoms of large spin $J=8$ \cite{chalopin2018quantum}. 
On the other hand, quadratic shifts are often detrimental for precise measurements, such as in atomic clocks. 
Even a small energy shift can be a dominant uncertainty in state-of-the-art precise measurements. 
Quadratic shifts are also harmful in spin detection via Faraday rotation \cite{Isayama1999,PhysRevLett.93.163602}.
%\cite{kuzmich1998atomic, thomsen2002spin, thomsen2002continuous, silberfarb2003continuous}.
In a Faraday rotation measurement, 
near-resonant light gives a large signal but may also change the atomic spin state
through nonlinear spin evolution due to quadratic shifts \cite{PhysRevLett.93.163602, colangelo2013quantum, Deutsch2010}.

Accurate measurement of quadratic or tensor shifts is important for building a sound basis for quantum control as well as for precise measurements. 
The tensor shifts in clock transitions in alkali atoms have been measured using 
the Ramsey method with a hot vapor \cite{Levi2016} and a cold-atom fountain \cite{costanzo2016ac}. 
Tensor shifts in cold lanthanide atoms in the ground state have been determined 
using Kapitza--Dirac diffraction from a pulsed standing wave \cite{kao2017anisotropic}, trap frequency measurement \cite{becher2018anisotropic, Ravensbergen2018} and modulation spectroscopy in an optical lattice \cite{Kreyer2021}. 
In these measurements \cite{Levi2016, costanzo2016ac, kao2017anisotropic, becher2018anisotropic, Ravensbergen2018, Kreyer2021}, 
the tensor shift is distinguished from other shifts based on the difference in the frequency dependence of the scalar and tensor shifts and/or the polarization dependence of the tensor shift.
%using several experimental data with different parameters such as light frequencies and polarization states. 
Tensor shift measurements distinguished in this way tend to be uncertain due to technical issues, such as imperfect polarization control at the atomic position.
It is difficult to precisely determine a tensor shift much smaller than a state-independent scalar shift.
This is the case for most atom experiments, 
although lanthanide atoms can have large tensor polarizability at specific light frequencies
\cite{kao2017anisotropic, becher2018anisotropic, Ravensbergen2018, Kreyer2021}.

In the present study, we demonstrate the detection of quadratic light shifts using a multi-state atomic interferometer \cite{Petrovic_2013, Sadgrove2013} in a Bose--Einstein condensate (BEC) of $^{87}$Rb atoms. 
This scheme is insensitive to state-independent light shifts and directly measures the quadratic light shift, thereby realizing a sensitive measurement 
by avoiding the uncertainty in distinguishing the tensor shift from other shifts.
Direct tensor shift detection is also advantageous in that we can measure the shift in a particular experimental configuration without needing to change light frequencies or polarization.
Furthermore, as we can measure the tensor shift without relying on a
priori theoretical knowledge including light frequency and polarization
dependence of the shift,
the measurement may be used for checking the validity of a theory.
Conversely, by verifying that the measured frequency dependence of the
quadratic shift is consistent with the theory, the validity of the
measurement scheme is confirmed.
We confirm the validity of our measurement in this manner.  
We also demonstrate suppression of nonlinear spin evolution using dual-color light pulses near the D$_1$ transition, with the light frequencies and powers chosen on the basis of the light shift measurement
to null the net quadratic shift.

The paper is organized as follows.
In Sec. \ref{sec: method} we present our experimental method and setup.
The experimental results are described in Sec. \ref{sec: result}.
We conclude the paper in Sec. \ref{sec: conclusion}.

\begin{figure*}
    \centering
    \includegraphics[width=1\linewidth]{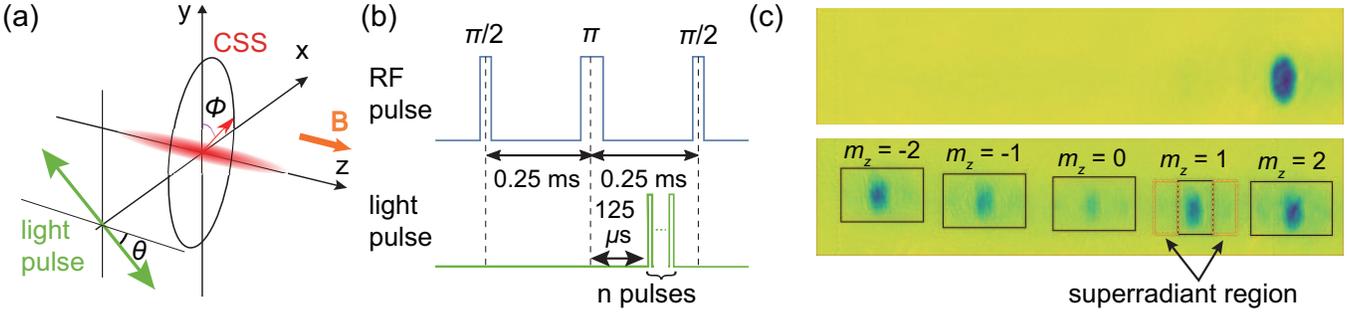}
    \caption{(color online) 
    (a) Experimental configuration.
    The BEC in a coherent spin state exhibits Larmor precession about the $z$ axis along the bias magnetic field, $B$.
    A linearly polarized 795-nm light pulse propagates along the $x$ direction through the BEC. 
   (b) Time sequence for quadratic shift detection.
   We use $\pi/2$-$\pi$-$\pi/2$ RF pulses to construct an atomic interferometer. 
   A train of $n$ light pulses is applied between the $\pi$ pulse and last $\pi/2$ pulse. 
   (c) Typical TOF images of a BEC measured after the detection sequence.
   The field of view is $2.1$ mm $\times$ $0.46$ mm.
   The upper and lower panels show TOF images without light pulses 
   and with a train of $n=9$ pulses of $P=5.71$ mW and $\Delta/2\pi = -840$ MHz,
   respectively.
   The black rectangles in the lower panel indicate the analysis regions for counting the atom number in each magnetic sublevel.
   The regions indicated by the red dotted boxes are used for counting atoms scattered by the superradiance in the $m_z = 1$ state.
   }
   \label{fig1}
\end{figure*}

\section{Experimental method and setup}\label{sec: method}
We produce a BEC of typically $3 \times 10^5 $ $^{87}$Rb atoms in a vacuum glass cell \cite{sekiguchi2020sensitive}.
The BEC is trapped in a crossed optical trap.
The trap is composed of an axial beam at a wavelength of 852 nm and a radial beam at 976 nm.
The axial and radial beam waists are approximately 30 $\mu$m and 70 $\mu$m, respectively. 
The axial and radial trap frequencies are measured to be $2 \pi \times (123, 16)$ Hz.
A bias magnetic field $B$ of 15 $\mu$T is applied along the axial beam along the $z$ axis.
We initially prepare the atoms in the $|F,m_z \rangle = |2,2 \rangle$ state, where $F$ is the quantum number for the total angular momentum of the atoms in the ground state
and $m_z$ denotes the magnetic sublevel.

The experimental configuration for the light shift measurement is shown in Fig.~\ref{fig1}(a).
We measure the quadratic light shift due to a light pulse near-resonant to the D$_1$ line ($\lambda = 795$ nm),
which propagates along the $x$ direction.
The D$_1$ light is generated by a distributed feedback (DFB) laser 
and is frequency offset-locked to a master external cavity diode laser (ECDL).
The frequency of the ECDL is stabilized to the $F = 2 \rightarrow F' = 1$ resonance line,
where $F'$ represents the total angular momentum for the excited $^2$P$_{1/2}$ state.
The beam power is controlled by an acousto-optic modulator (AOM).
The beam is directed to the atoms through an optical fiber after the AOM.
The beam is almost collimated before the atom cell 
and has a Gaussian profile with a $1/e^2$-radius of $\omega _0 =0.75$ mm.
We control the polarization state at the atomic position using a half-wave plate (HWP) and a quarter wave plate (QWP) before the atom cell.
Just after the cell we adjust the polarization of the light to be linearly polarized.
The angle between the polarization plane and the direction of the magnetic field, $\theta$, is adjusted to 54.7$^{\circ}$, 
at which the nonlinear spin evolution of the precessing atoms is averaged over a Larmor period if the light is continuous \cite{PhysRevLett.93.163602, sekiguchi2020sensitive}.

\begin{figure*}
    \centering
   \includegraphics[width=1\linewidth]{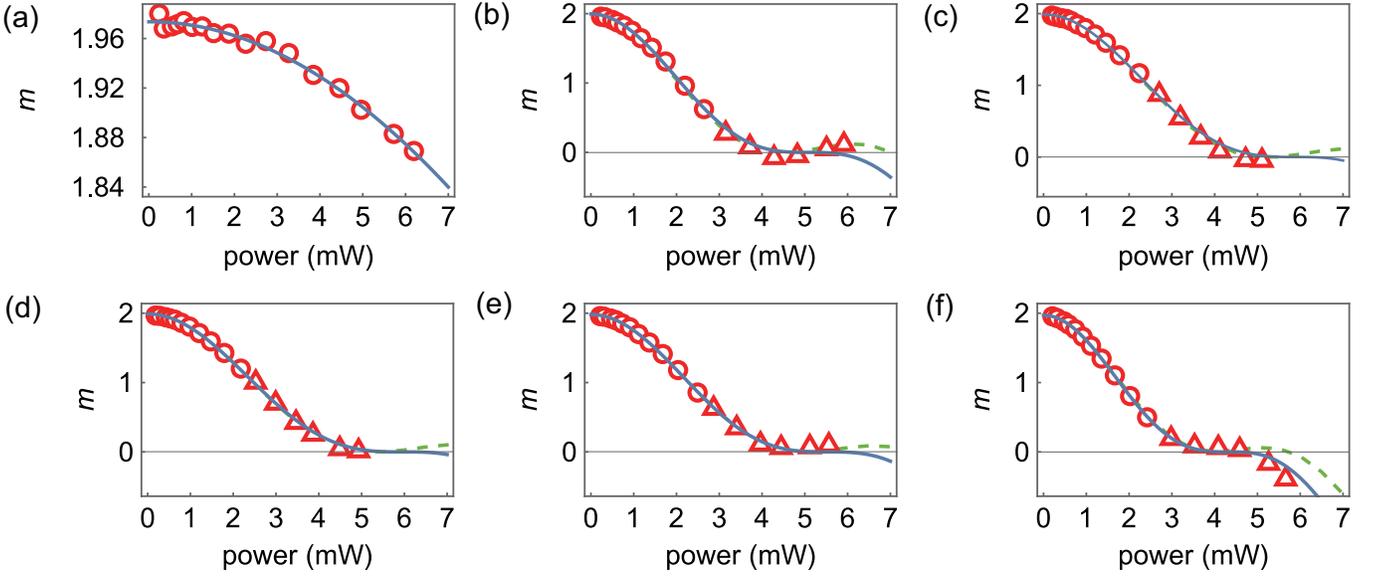}
    \caption{(color online) Power dependence of magnetization, $m$,
    for $\Delta / 2\pi$ of (a) $-840$ MHz,  (b) 240 MHz, (c) 340 MHz, (d) 440 MHz, (e) 540 MHz and (f) 640 MHz.   
   The red symbols (circles and triangles) represent the measured magnetization. 
   The circle data points are used for fitting (see text for details)
   and the blue solid lines are the fitting curves.
   The green dashed line is the simulated magnetization including the spontaneous emission effect.
   The calculation and simulation are performed for a single pulse of $\tau=667$ ns.
   }
   \label{fig2}
\end{figure*}

We perform atomic interferometry with the time sequence depicted in Fig.~\ref{fig1}(b).
In this sequence, we use the spin echo method to suppress the influence from low-frequency fluctuations of the bias magnetic field. 
We apply light pulses between the middle $\pi$ pulse and last $\pi / 2$ rf pulse. 
Each magnetic sublevel experiences an AC Stark shift and acquires a phase shift during the light pulse.
Different phase shifts between the magnetic sublevels result in a change in the population of each sublevel after the last $\pi /2$ rf pulse.
We measure the sublevel populations by absorption imaging along the $x$ axis after a time-of-flight of $20.6$ ms with Stern--Gerlach spin separation (see Fig.~\ref{fig1}(c)).

The AC Stark shift due to light near-resonant to the D$_1$ line is derived from the light-shift Hamiltonian \cite{Deutsch2010}:
%\begin{eqnarray}
%    \hat{H}_{\mathrm{atom}} &\simeq& \sum_{f'} \frac{Ap}{\Delta_{ff'}} \{ C_{ff'}^{(0)} |\vec{\varepsilon}|^2 + \mathrm{i}C_{ff'}^{(1)} (\vec{\varepsilon}^{\, *} \times \vec{\varepsilon}) \cdot \hat{\bm{F}} \nonumber \\
%    &+& C_{ff'}^{(2)} (|\vec{\varepsilon} \cdot \hat{\bm{F}}|^2 - \frac{1}{3} \hat{\bm{F}}^2 |\vec{\varepsilon}|^2) \} ,
%\end{eqnarray}
\begin{eqnarray}
    \hat{H}_{\mathrm{shift}} &=& \sum_{F'} \frac{\hbar\Omega_0^2}{4\Delta_{FF'}} \{ C_{FF'}^{(0)} |\vec{\varepsilon}|^2 + \mathrm{i}C_{FF'}^{(1)} (\vec{\varepsilon}^{\, *} \times \vec{\varepsilon}) \cdot \hat{\bm{F}} \nonumber \\
    &+& C_{FF'}^{(2)} (|\vec{\varepsilon} \cdot \hat{\bm{F}}|^2 - \frac{1}{3} \hat{\bm{F}}^2 |\vec{\varepsilon}|^2) \} ,
\end{eqnarray}
where $\Delta_{FF'}$ is the amount of light detuning from the transition frequency between the $F$ and $F'$ states,
$C_{FF'}^{(k)}$ is a rank-$k$ tensor coefficient representing the angular momentum dependence \cite{Deutsch2010}
and $\vec{\varepsilon}$ is the polarization vector for the light. 
%\begin{eqnarray}
 %   A &=& \frac{| \langle P_{1/2} || d || S_{1/2} \rangle |^2 g}{\hbar}, \\ 
 %   g &=& \frac{1}{\pi r^2 c \epsilon_0} ,
%\end{eqnarray}
$\Omega_0$ is defined by
\begin{equation}
\Omega_0=\frac{ \langle P_{1/2} || d || S_{1/2} \rangle E}{\hbar},
\end{equation}
where $\langle P_{1/2} || d || S_{1/2} \rangle = 2.537 \times 10^{-29} \, \mathrm{C} \cdot \mathrm{m}$ \cite{loudon2000quantum} is the reduced matrix element for the D$_1$ dipole transition and $E$ is the field amplitude.
$\Omega_0$ can be expressed using the beam power, $P$, as 
\begin{equation}
\Omega_0= \frac{ \langle P_{1/2} || d || S_{1/2} \rangle }{\hbar} \sqrt{ \frac{4P}{\pi c \varepsilon_0 \omega_0^2} } \equiv \sqrt{\eta P}, 
\end{equation}
where $c$ is the speed of light and  $\varepsilon_0$ is the electric  constant.
%$C_{ff'}^{(k)}$ is a rank-$k$ tensor coefficient represented by Wigner--6j symbol, $\langle P_{1/2} || d || S_{1/2} \rangle$ is a reduced matrix element for D1 line, 
%$\vec{\varepsilon}$ is a polarization vector, $p$ is a light power and $r$ is a beam radius of laser.
% assume that polarization is linear, we can neglect a vector shift.
Hereafter we consider linearly polarized light,
which introduces no vector shift and produces a state-dependent shift solely through the tensor component.
The state-dependent Hamiltonian can be written as
\begin{equation}
    \hat{H}_{\mathrm{depend}} = \frac{\hbar\Omega_0^2}{4\Delta_{\mathrm{HFS}}} \chi (\hat{F_y} \mathrm{cos} \theta + \hat{F_z} \mathrm{sin} \theta)^2,
\end{equation}
% \begin{equation}
%     \hat{H}_{\mathrm{depend}} = \sum_{f'} \left( \frac{\hbar\Omega_0}{2} \right)^2 a (\hat{F_y} \mathrm{cos} \theta + \hat{F_z} \mathrm{sin} \theta)^2,
% \end{equation}
%\begin{equation}
%    \hat{H}_{\mathrm{depend}} = \sum_{f'} A \hbar a p (\hat{F_y} \mathrm{cos} \theta + \hat{F_z} \mathrm{sin} \theta)^2 .
%\end{equation}
where $\Delta_{\mathrm{HFS}} = 2 \pi \times 814.5$ MHz is the hyperfine splitting between the $F'=1$ and $F'=2$ states and
\begin{equation}
    \chi = \sum_{F'} \frac{ C_{FF'}^{(2)} \Delta_{\mathrm{HFS}}} {\Delta_{FF'}}
\end{equation}
% \begin{equation}
%     a = \sum_{f'} \frac{ C_{ff'}^{(2)} } { \hbar \Delta_{ff'}}
% \end{equation}
represents the dependence of the coupling strength on the light frequency.
We refer to $\chi$ as the coupling coefficient.
In an experiment using the ground $F=2$ state, $\chi$ depends on the laser frequency as
\begin{equation}\label{eq:a}
    \chi = \frac{\Delta_{\mathrm{HFS}}}{12} \left( \frac{1}{\Delta} - \frac{1}{\Delta - \Delta_{\mathrm{HFS}}} \right),
\end{equation}
where $\Delta \equiv \Delta_{21}$ (see Fig.~\ref{fig4}(b)).
Hereafter, we consider detuning with respect to the transition frequency between the $F=2$ and $F'=1$ states.
%$\theta$ is the polarization angle from $y$ axis and $a = \frac{1}{\hbar} \sum_{f'} C_{f'f}^{(2)} / \Delta_{ff'}$ is a coefficient representing the nonlinear phase shift during the pulses depending on detuning.

The spin dynamics for $\hat{H}_{\mathrm{depend}}$ can be easily described if the quantization axis is selected to be along the direction of the polarization axis, $z'$.
The state before sending the light pulse, $|\psi \rangle = \sum_{m_{z'}} \beta_{m_{z'}} |m_{z'} \rangle$, 
with $\beta_{m_{z'}}$ being the probability amplitude for each sublevel $m_{z'}$ in this frame,
evolves under the influence of a rectangular pulse of width $\tau$ and power $P$ into
\begin{equation}\label{eq: evolution}
    |\psi'\rangle = \sum_{m_{z'}} \beta_{m_{z'}} e^{\mathrm{i} \chi {m_{z'}}^2 \xi P \tau}|m_{z'}\rangle,
\end{equation}
%\begin{equation}
%    |\psi'\rangle = \sum \beta_m e^{\mathrm{i}\tau Aapm^2}|m\rangle
%\end{equation}
where
\begin{equation}
    \xi = \frac{\eta}{4\Delta_{\mathrm{HFS}}}.
\end{equation}
Here we consider evolution due to a single pulse for simplicity. The extension to the multi-pulse case is straightforward.
We assume that the pulse width, $\tau$, is sufficiently shorter than the period of Larmor precession and
neglect evolution due to the magnetic field during the pulses.
In Eq.~(\ref{eq: evolution}) we omit the global (state-independent) phase shift, which has no relevance to the population change in the magnetic sublevels.
If the spin evolution during the pulse is purely caused by the light shift,
the state after the last $\pi/2$ pulse in the $m_z$ basis is written
using the Wigner D-matrix \cite{Sadgrove2013}, $D^j (\alpha,\beta,\gamma)$, as
\begin{eqnarray}\label{eq:phiend}
    | \psi _{\mathrm{end}} \rangle &=& D^2 (0,-\frac{\pi}{2},0) D^2 (0,0,-\phi) D^2 (-\theta,0,0)^{\dagger} \nonumber \\
    &&A D^2 (-\theta,0,0) D^2 (0,0,\phi) | \psi  _0 \rangle,
\end{eqnarray}
% \begin{eqnarray}\label{eq:phiend}
%     | \psi \rangle _{\mathrm{end}} &=& D^2 (\pi,0,0) D^2 (0,-\frac{\pi}{2},0) \nonumber \\
%     &&D^2 (0,-\theta,\phi)^{\dagger} A D^2 (0,-\theta,\phi) | \psi \rangle _0,
% \end{eqnarray}
where
%\begin{equation} 
$A = \mathrm{diag}(e^{4\mathrm{i} \chi \xi P \tau},e^{\mathrm{i} \chi \xi P \tau},1,e^{\mathrm{i} \chi \xi P \tau},e^{4\mathrm{i} \chi \xi P \tau})$ represents the time development by light pulses,
%\end{equation}
%\begin{equation}
%U = D^2 (0,-\theta,\phi),
%\end{equation}
$\phi$ is the spin angle in the $x$--$y$ plane with respect to the $y$ axis
at the starting time of the light pulse
and $|\psi _{0} \rangle = (-1/4,\mathrm{i}/2,\sqrt{6}/4,-\mathrm{i}/2,-1/4)^T$ represents the coherent spin state (CSS) along the $y$ axis ($\phi=0$) as the basis of the $z$ axis.
Here $D^j (\alpha,\beta,\gamma) $ is defined in terms of the Euler angles $(\alpha,\beta,\gamma)$ around the $z$--$y$--$z$ axes as
\begin{equation}
    D_{q'q}^j (\alpha,\beta,\gamma) = \langle jq' | \hat{R}(\alpha,\beta,\gamma) | jq\rangle,
\end{equation}
where $\hat{R}(\alpha,\beta,\gamma) = e^{-\mathrm{i}\alpha \hat{j}_z} e^{-\mathrm{i}\beta \hat{j}_y} e^{-\mathrm{i}\gamma \hat{j}_z}$ is the rotational operator \cite{sakurai_napolitano_2017}.
We calculate the magnetization, $m = \langle \psi _{\mathrm{end}} | \hat{F}_z | \psi _{\mathrm{end}} \rangle $, using Eq.~(\ref{eq:phiend}).
The calculated $m$ is a function of $\chi \xi P \tau$, but its explicit expression is too long to show here.
We note $m = 2\mathrm{cos}^3(\chi \xi P \tau)$ if $\theta = 0$ \cite{kitagawa1993squeezed}.

\section{Results}\label{sec: result}

\subsection{Measurement of quadratic light shifts}
We first detect quadratic light shifts produced by a single pulse of $\Delta/(2\pi) = -840$ MHz,
which induces less light-assisted collisional atom losses \cite{sekiguchi2020sensitive}.
The pulse has an almost rectangular shape and its
length is fixed to $\tau = 667 \,\mathrm{ns}$.
The pulse is applied $0.125 \,\mathrm{ms}$ after the $\pi$ pulse, when the spin orientation is along the $x$ axis ($\phi=\pi/2$).
We confirm the spin direction by spin-sensitive phase contrast imaging \cite{sekiguchi2020sensitive}.

We observe changes in the sublevel population using a Stern--Gerlach measurement,
as shown in Fig.~\ref{fig1}(c).
We experimentally obtain the magnetization using
\begin{equation}
m = \frac{\sum_i i N_i}{N_{\mathrm{tot}}},
\end{equation}
where $N_i$ is the number of atoms in the $|F, m_z=i \rangle$ state $(i=-2,-1,0,1,2)$ after the read-out pulse and $N_{\mathrm{tot}}=\sum_i N_i$ is the total number of atoms.
The magnetization is plotted as a function of the pulse power in Fig.~\ref{fig2} (a).
We fit the data using $(1-\delta p) \langle \psi _{\mathrm{end}} | \hat{F}_z | \psi _{\mathrm{end}} \rangle$,
where $\delta p$ is introduced to account for experimental imperfect state preparation and control.
In the fitting (blue line in Fig.2 (a)), we use Eq.~(\ref{eq:phiend}) with $\theta = 54.7^{\circ}$ and $\phi = 90^{\circ}$, 
in correspondence with the experiment.
The fitting gives $\chi = -0.0404(8)$.
The value in parentheses denotes the standard deviation of $\chi$ calculated from three sets of data.

Next, we measure the quadratic light shifts for other light frequencies with positive detunings $\Delta/(2\pi) = +\{ 240, 340, 440, 540, 640 \}$ MHz. 
The results for each detuning are depicted in Figs.~\ref{fig2}(b)--\ref{fig2}(f).
We observe large changes in $m$.
For these positive detunings, 
$\chi$ is much larger than that for $\Delta/(2\pi) = -840$ MHz and 
the observed large change in $m$ is reasonable.
The spontaneous emission rates also become large for these detunings and
optical pumping due to spontaneous emission may also result in spin change.
To evaluate the contribution of spontaneous emissions,
we numerically calculate the dynamics of the magnetization using an atomic master equation
including spontaneous emission \cite{happer2010optically}.
A theoretically predicted value of the coupling coefficient, $\chi_{\textrm{theo}}$, is used in the simulation. 
We plot the simulation results in Fig.~\ref{fig2} (green dotted lines).
The numerical simulation indicates that the spontaneous emissions are not negligible 
for beam powers larger than approximately 5 mW, which we were not able to investigate in detail due to the limited available beam power.
In the fitting shown by blue lines in Fig. 2(b)--(f), to obtain experimental values of the coupling coefficient, $\chi_{\textrm{exp}}$, 
we use data points at beam powers for which
the magnetization obtained by the simulation differs from the matrix calculation by less than 0.06
to lessen the effect of spontaneous emissions on the estimation of $\chi$.
The fitted values of $\chi_{\textrm{exp}}$ are shown in Table \ref{tab1}.

\begin{figure}    \centering
   \includegraphics[width=8.6cm]{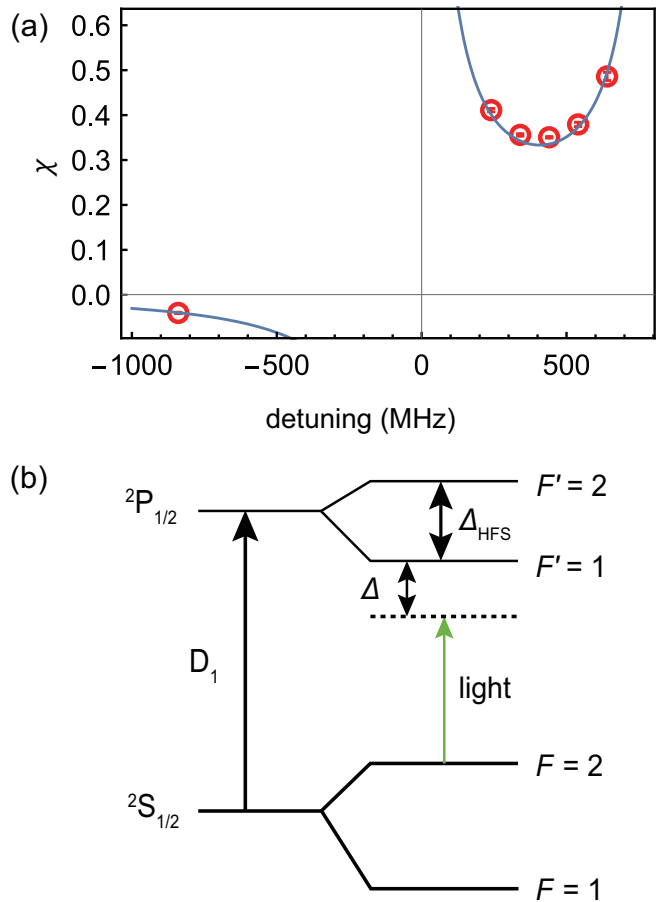}
    \caption{(color online) (a) Dependence of the coupling coefficient $\chi$ on detuning.
   The red circles represent the coefficients obtained from the experiments.
   The error bar for each circle represents the standard deviation over three experimental runs.
   The blue solid line is the theoretical curve given by Eq.~(\ref{eq:a}).
   (b) Energy diagram for $^{87}$Rb relevant to the experiment.
   }
   \label{fig4}
\end{figure}

The frequency dependence of $\chi_{\textrm{exp}}$ is consistent with the theoretical curve given by Eq.~(\ref{eq:a}), as shown in Fig.~\ref{fig4}(a).
Note that interferometric detection does not reveal the sign of $\chi$.
We determine the sign of $\chi$ at each $\Delta$ in Fig.~\ref{fig4}(a) to coincide with the theoretically determined sign.
We also note the exact determination of $\chi$ requires precise calibration of the beam intensity at the atom position.
It is more appropriate to analyze the ratio between $\chi_{\textrm{exp}}$ and $\chi_{\textrm{theo}}$ to estimate the validity of the measurement.
This ratio is shown in Table \ref{tab1}.
The sample standard deviation of the ratios, which represents the precision of the measurement, is $0.02$ (2\%).

\subsection{Suppression of nonlinear spin evolution}

We also demonstrate suppression of spin evolution due to a quadratic shift by using near-resonant dual-color light pulses. 
From Eq.~(\ref{eq:a}), we can see that $\chi$ is positive if $0<\Delta <\Delta_{\mathrm{HFS}}$ and negative otherwise.
%has opposite sign depending on whether detuning is positive ($0 < \Delta < \Delta_{21}$) or negative.
%If detuning is positive ($0 < \Delta < \Delta_{21}$), $a$ has positive value, and vice versa.
If we combine negative ($\Delta<0$) and positive ($0<\Delta<\Delta_{\mathrm{HFS}}$) detuned light with a power ratio of $p_{\mathrm{neg}} / p_{\mathrm{pos}} = \chi_{\mathrm{pos}} / \chi_{\mathrm{neg}}$,
the quadratic light shift should vanish and nonlinear spin evolution should be suppressed. 
%enables to vanish the overall tensor Hamiltonian and suppress the nonlinear phase shifts
We use the DFB laser used in the above experiment at a fixed detuning of $\Delta_-/(2\pi)=-840$ MHz.
We prepare another ECDL for quadratic shift compensation.
The ECDL is frequency offset-locked to the master ECDL, and its detuning is set to $\Delta_+/(2\pi) = +\{ 240, 340, 440, 540, 640 \}$ MHz.
The two laser beams are mixed at a non-polarizing beam splitter (NPBS) before the AOM for power control.
We adjust the ratio of the DFB laser power to the ECDL power using a HWP and PBS before the NPBS.
%control the intensity of positive detuned beam by using a HWP and a polarization beam splitter (PBS) 
%before entering the NPBS
%to adjust the ratio between the positive and negative beams.
%We mixed two beams of positive and negative detunings.
The ratios used in the experiment, $r_{\mathrm{exp}}$, are listed in Table.~\ref{tab1}.
The values of  $r_{\mathrm{exp}}$ are close to the theoretical optimal ratios, 
\begin{equation}
    r_{\mathrm{theo}} = \left( \frac{1}{\Delta_+} - \frac{1}{\Delta_{+} - \Delta_{\mathrm{HFS}}} \right) / \left( \frac{1}{\Delta_-} - \frac{1}{\Delta_{-} - \Delta_{\mathrm{HFS}}} \right).
\end{equation}
%  $r_{\mathrm{theo}} = \left( \frac{1}{\Delta_+} - \frac{1}{\Delta_{+} - \Delta_{\mathrm{HFS}}} \right) / \left( \frac{1}{\Delta_-} - \frac{1}{\Delta_{-} - \Delta_{\mathrm{HFS}}} \right)$.
%$r_{\mathrm{exp}}$ is optimal ratio derived by the result of experiment A, and $r_{\mathrm{theo}} = \left( \frac{1}{\Delta} - \frac{1}{\Delta_{21} - \Delta} \right) / \left( \frac{1}{\Delta_{\mathrm{probe}}} - \frac{1}{\Delta_{21} - \Delta_{\mathrm{probe}}} \right)$ 
%is theoritical optimal ratio where $\Delta_{\mathrm{probe}} = -840$ MHz.

\begin{table}[htb]
  \centering
    \caption{Measured and theoretical coupling coefficients and power ratios for quadratic shift cancellation.}
    \begin{tabular}{cccccc} \hline \hline
      $\frac{\Delta}{2\pi}$ (MHz) & $\chi_{\mathrm{exp}}$ & $\chi_{\mathrm{exp}}/\chi_{\mathrm{theo}}$&$r_{\mathrm{exp}}$ & $r_{\mathrm{theo}}$ & $r_{\mathrm{exp}} / r_{\mathrm{theo}}$ \\ \hline 
      $-840$ & $-0.0404$ & 1.016 &$- $& $-$ & $-$\\ 
      240 & 0.411 & 1.025 &10.17 & 10.08 & 1.01\\
      340 & 0.356 & 1.039 &8.81 & 8.61 & 1.02\\
      440 & 0.351 & 1.045 &8.68 & 8.43 & 1.03\\
      540 & 0.379 & 1.017 &9.39 & 9.37 & 1.00\\
      640 & 0.486 & 0.982 &12.02 & 12.44 & 0.97\\ \hline \hline
    \end{tabular}
    \label{tab1}
  \end{table}

\begin{figure}
    \centering
    \includegraphics[width=8.6cm]{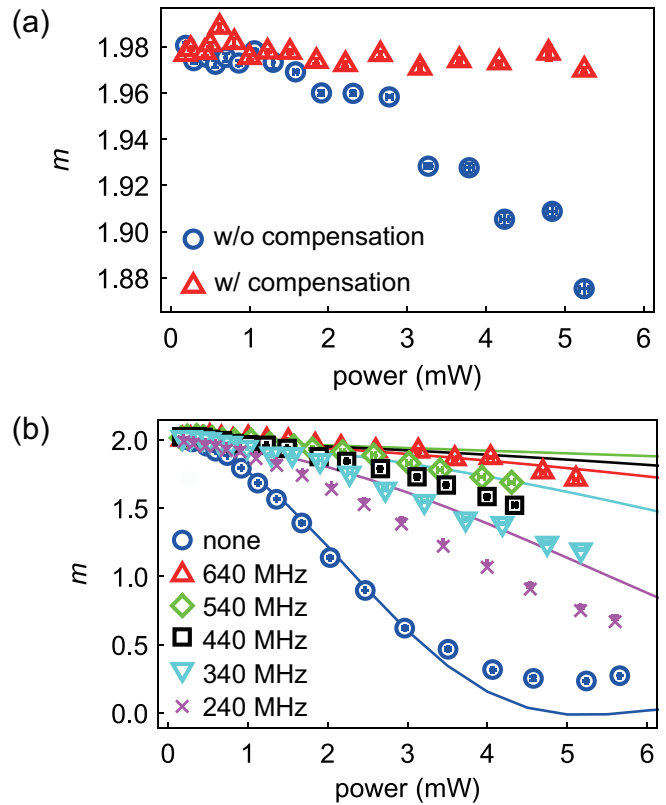}
     \caption{(color online) Magnetization with and without compensation as function of negative-detuned beam power.
     (a) Magnetization for single pulse. 
     The blue circles represent the magnetization without compensation (with only the negative-detuned light).
     The red triangle data are taken with the compensation light with $\Delta_+/(2\pi)=+640$ MHz added.
     (b) Magnetization for $n = 9$ pulses.
     The magnetization values with compensation light of $\Delta_+/(2\pi)=$ 240, 340, 440, 540, 640 MHz 
     are plotted by pink crosses, light blue inverted triangles, black squares, green diamonds and red triangles, respectively.
     The blue circles represent the magnetization without compensation.
     The solid lines are the magnetization by the numerical simulation including spontaneous emissions.
     % of atomic master equation, which include spin relaxation by spontaneous emissions. 
    }
    \label{fig5}
\end{figure}

\begin{figure}
    \centering
   \includegraphics[width=8.6cm]{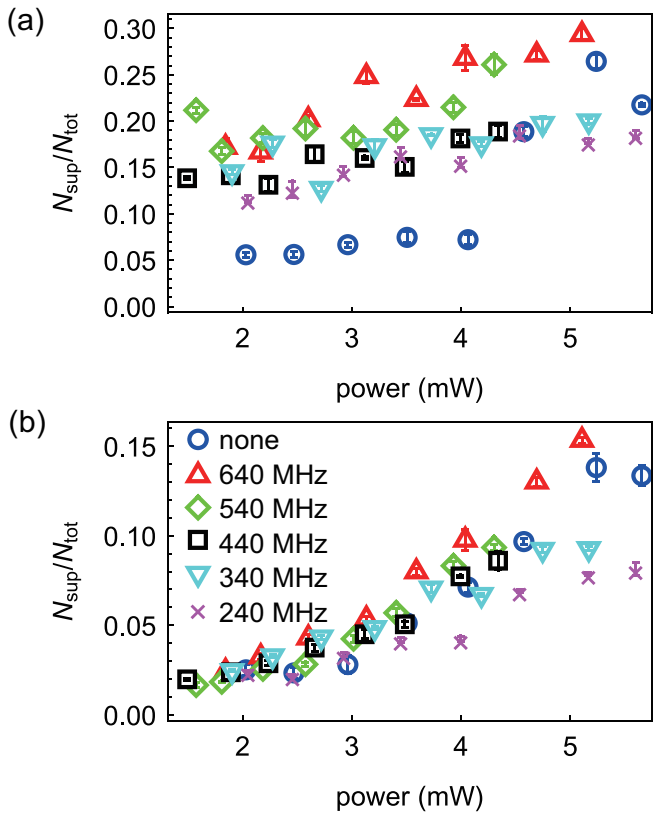}
    \caption{(color online) Proportion of atoms kicked by superradiance
    for (a) $m_z=+1$ and (b) $m_z=+2$ as function of negative-detuned beam power.
   }
   \label{fig6}
\end{figure}

We observe that the change in magnetization for the pulse mixture 
is much less than that for the single negative-detuned pulse.
We show the results for $\Delta_+ /(2\pi) = +640$ MHz light compensation and 
without compensation in Fig.~\ref{fig5}(a).
We apply $n = 9$ pulses to observe the changes in magnetization more clearly.
We set the interval between pulses to $9.5 \,\mu$s, equal to the period of Larmor precession, 
in order to apply all pulses when the spin is directed to the probe propagation direction.
The results with compensation light pulses with $\Delta_+/(2\pi)=$ 240, 340, 440, 540, 640 MHz are 
plotted in Fig.~\ref{fig5}(b).

\begin{figure}
    \centering
   \includegraphics[width=8.6cm]{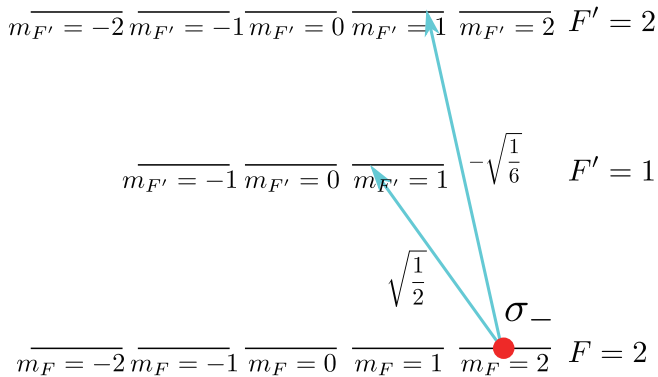}
    \caption{(color online) 
    Coupling strength between $|F, m_F\rangle=|2,2\rangle$ state and excited states.
    The $x$ axis is taken as the quantization axis here.
    The blue arrows represent the allowed transitions.
    The number next to each arrow is the dipole matrix element, expressed as multiples of $\langle P_{1/2} || d || S_{1/2} \rangle$.
   }
   \label{fig7}
\end{figure}

The magnetization changes with increasing power even with compensation.
This change is considered to be due to spin relaxation by optical pumping.
We also observe that the size of the change depends on the detuning of the compensation beam, $\Delta_+$.
The closer the frequency is to the $F=2 \rightarrow F'=1$ resonance, the greater the magnetization change.
The frequency dependence can be understood as follows.
When we take the $x$ axis as the quantization axis, the linearly polarized beam couples to the atoms 
via the circular component, as shown in Fig.~\ref{fig7}.
The coupling strength to the $F'=1$ state is three times stronger than that to the $F'=2$ state. 
The increase in the magnetization change as $\Delta$ approaches 0 is related to this difference 
in coupling strength.
The frequency dependence of the magnetization change is confirmed by a numerical simulation.
The simulated change is minimum at $\Delta_+/(2\pi)=$ 540 MHz.
This is consistent with the simple estimation of the relaxation rate
given by the product of the scattering rate (approximately proportional to $3/\Delta_{21}^2 + 1/\Delta_{22}^2$) and the beam power required for the quadratic shift compensation ($\propto 1/\Delta_{21} - 1/\Delta_{22}$).

The magnetization change in the experiment is larger than that in the simulation including the spontaneous emissions,
especially for detuning from 340 to 540 MHz.
A possible cause of this difference is the superradiance \cite{PhysRev.93.99, gross1982superradiance, rehler1971superradiance}.
In fact, we observe atoms kicked by superradiant scattering when the beam is strong.
We evaluate the proportion of the number of kicked atoms, $N_{\mathrm{sup}}$, to
the total number of atoms, $N_{\mathrm{tot}}$, as a measure of superradiant scattering.
In the analysis, we count the number of kicked atoms in the regions adjacent to the zero moment component (see the lower panel of Fig.~\ref{fig1}(c)) as $N_{\textrm{sup}}$.
We plot $N_{\mathrm{sup}}/N_{\mathrm{tot}}$ for the magnetic sublevels of $m_z=+1$ and $+2$ in Fig.~\ref{fig6}.
The data in Fig.~\ref{fig6} is taken with $n=9$ pulses.
We consider that superradiance enhances the spin relaxation rate 
and may increase the magnetization change.
We estimate from Fig.~\ref{fig6}(a) that the scattering rate is enhanced by up to approximately $30\%$. 
The Raman superradiance \cite{Schneble2004} to the $F=1$ state should also contribute to the scattering rate enhancement.
We also measure the atom number loss, which changes the population among the magnetic sublevels and causes spin relaxation.
However, we do not observe a significant dependence of the loss rates on $\Delta_+$. 
The larger than expected magnetization change might be partly due to the enhancement of spontaneous emission near the critical temperature for Bose--Einstein condensation \cite{bons2016quantum}.

\section{Conclusions and outlook}\label{sec: conclusion}
We have demonstrated the successful measurement of quadratic light shifts using an atom interferometer.
The measured dependence of the shift due to detuning is consistent with the theoretical prediction.
In addition, we suppressed the influence of nonlinear light shifts by using dual-color light pulses.
Dual-color light is applicable to high-precision measurement in a probe, especially for BEC magnetometers \cite{sekiguchi2020sensitive}.
A dual-color probe is useful not only for improving the signal-to-noise ratio in spin measurements by increasing the probe strength, but also for preventing disturbances when creating a spin squeezed state
via quantum nondemolition measurements \cite{kuzmich1998atomic, thomsen2002spin, thomsen2002continuous, silberfarb2003continuous, PhysRevA.104.023710}.

Further improvements of this measurement scheme are feasible.
Reducing technical noises  during absorption imaging and spin manipulation by rf pulses will lead to higher sensitivity.
The bias magnetic field may produce a slight difference between the experimental data and the theoretical prediction calculated by the Wigner D-matrix and interaction Hamiltonian, in which we have neglected the spin evolution due to the magnetic field.
Therefore, to improve the estimation accuracy for the coupling coefficients, 
it may be effective to use a low magnetic field.
These improvements may make it possible to estimate physical quantities such as transition matrix elements and hyperfine splittings. 

\begin{acknowledgments}
This work was supported by the MEXT Quantum Leap Flagship Program (MEXT Q-LEAP) grant number JPMXS0118070326
and JSPS KAKENHI grant number JP19K14635.
\end{acknowledgments}

%\bibliography{ref}
% \bibliographystyle{unsrt}

\end{document}